\documentclass[preprint,prb]{revtex4}
\usepackage{amsmath}
\newcommand{{\bff}}{\mbox{\boldmath$f$\unboldmath}}

\newcommand{{\bfF}}{\mbox{\boldmath$F$\unboldmath}}

\newcommand{{\bfA}}{\mbox{\boldmath$A$\unboldmath}}

\newcommand{\gradv}{{\nabla}}

\def\v#1{{\bf#1}}

\begin{document}

\title{Comment on ``A generalized Helmholtz theorem for time-varying vector
fields,'' by Artice M. Davis [Am. J. Phys. {\bf 74}, 72--76 (2006)]}
\author{Jos\'e A. Heras}
\email{heras@phys.lsu.edu}
\affiliation{Departamento de F\'\i sica, E. S. F. M., Instituto
Polit\'ecnico Nacional, M\'exico D. F. M\'exico and 
Department of Physics and Astronomy, Louisiana State University, Baton
Rouge, Louisiana 70803-4001, USA}


\maketitle

In a recent paper Davis formulated the following generalization of
the Helmholtz theorem for a time-varying vector field:\cite{1} 
\begin{equation}
\bfF= \frac{1}{c^2}\frac{\partial}{\partial t}\bigg(\gradv \phi +
\frac{\partial\bfA}{\partial t} \bigg) +\gradv \times(\gradv \times\bfA), \label{eq:davis}
\end{equation}
where $\phi$ and $\bfA$ are the Lorenz gauge retarded potentials.
The purposes of this comment are to point out that Davis's generalization
is a version of the generalization of the Helmholtz theorem
formulated some years ago by McQuistan\cite{2} and Jefimenko\cite{3} and more
recently by the present author\cite{4,5,6} and to show that Davis's
expression for the field $\bfF$ is also valid for potentials in gauges
other than the Lorenz gauge.

The generalized 
Helmholtz theorem states that a retarded
vector field vanishing at infinity can be written as\cite{4}
\begin{equation}
\bfF=- \gradv\!\int\!d^3 x'\,\frac{[\gradv '\cdot\bfF]}{4\pi
R}+ \gradv \times\!\int\!d^3 x'\,\frac{[\gradv'\times\bfF]}{4\pi R} +
\frac{1}{c^2}\frac{\partial}{\partial t}\!\int\!d^3 x'\,\frac{[\partial
\bfF/\partial t]}{4\pi R}, \label{eqtheor}
\end{equation}
where the square brackets denote the retardation symbol, $R=|\v x-\v x'|$,
and the integrals are over all space. If we define the
potentials
$\Phi$, $\v A$, and $\v C$ by
\begin{subequations}
\begin{align}
\Phi & =\!\int\!d^3 x'\,\frac{[\gradv'\cdot\bfF]}{4\pi R},\\
\v A & =\!\int\!d^3 x'\,\frac{[\gradv'\times\bfF]}{4\pi R},\\
\v C& =\!\int\!d^3 x'\,\frac{[\partial \bfF/\partial t]}{4\pi R}, 
\end{align}
\end{subequations}
then Eq.~(\ref{eqtheor}) can be written compactly as 
\begin{equation}
\bfF=-\gradv\Phi +\gradv\times\v A+ \frac{1}{c^2}\frac{\partial\v
C}{\partial t}.
\end{equation}
The potentials $\Phi$ and $\v A$ in this formulation of the theorem are
different from the potentials $\phi$ and $\bfA$ in Davis's formulation. It is
not difficult to derive the relations $\partial \phi/\partial
t=-c^2\Phi$, $\gradv\times\bfA=\v A$, and $\partial\bfA/\partial t=\v C,$
which imply the formal equivalence between the two formulations.

The standard
Helmholtz theorem is usually applied to solve the equations of
electrostatics and magnetostatics. The generalization of this
theorem\cite{4} can be used to solve Maxwell's equations. The generalization
proposed by Davis\cite{1} can be used to elucidate the form of Maxwell's equations. The two versions of the generalized Helmholtz theorem are complementary.

In Ref.~\onlinecite{1} 
the electric field
$-\gradv\phi - \partial\bfA/\partial t$ and the magnetic field
$\gradv\times\bfA$ are expressed in terms of the Lorenz gauge potentials, which were used to formulate Eq. (1) for the time-varying vector field $\bfF$. Equation~(\ref{eq:davis}) can also be formulated using potentials in other gauges. For example, it can
be formulated for potentials in the velocity gauge\cite{7}
$\gradv\cdot \bfA+(1/v^2)\partial \phi/\partial t=0$, a class of gauges
containing the Coulomb gauge $(v=\infty)$, the Lorentz gauge $(v=c)$, and
the Kirchhoff gauge\cite{8} $(v=ic)$. Jackson\cite{7}
recently derived the gauge function $\chi_v$ (Eq.~(7.5) of
Ref.~\onlinecite{7}), which transforms the
Lorenz gauge potentials $\phi_L$ and
$\bfA_L$ to the velocity gauge potentials $\phi_v$ and $\bfA_v$:
\begin{subequations}
\begin{align}
\phi_v & =\phi_L -\frac{\partial\chi_v}{\partial t}, \\
\bfA_v & =\bfA_L +\gradv\chi_v.
\end{align}
\end{subequations}
From Eqs. (5) we
obtain
\begin{subequations}
\label{equalities}
\begin{align}
\gradv\phi_L + \frac{\partial\bfA_L}{\partial t} & =\gradv\phi_v +
\frac{\partial\bfA_v}{\partial t} \\ 
\gradv\times \bfA_L & =\gradv\times \bfA_v.
\end{align}
\end{subequations}
Equations~\eqref{equalities} imply that Eq. (1) is also
valid for potentials in the velocity gauge, which means that it is
valid for the Coulomb and Kirchhoff gauges also. The application of Eq. (1) to potentials in the velocity gauge 
requires the identification $\bfF=\mu_0\v J$, where $\v J$ is the
current density and $\mu_0$ the permeability of free space.

Davis introduced causality in Eq.~\eqref{eq:davis}
when he chose retarded potentials. But we can equally choose acausal
advanced potentials to obtain Eq.~\eqref{eq:davis}. Causality in Eq. (1) is not a necessary assumption, but it is required to identify 
$-\gradv\phi - \partial\bfA/\partial t$ and $\gradv\times\bfA$ with the
retarded electric and magnetic fields. As pointed out by Rohrlich,\cite{9}
causality must be inserted by hand in classical field theories as a
condition.

The reader might wonder why Eq.~\eqref{eq:davis} can also
be written in terms of the Coulomb gauge potentials when the instantaneous
scalar potential $\phi_C$ in this gauge is clearly acausal. The explanation
is that the Coulomb gauge vector potential $\v A_C$ contains two parts,
one of which is causal (retarded) and the other is acausal
(instantaneous). Jackson\cite{7} recently derived a novel expression for
$\v A_C$ (Eq.~(3.10) in Ref.~\onlinecite{7}) which exhibits 
both parts. The fact that $\v A_C$ carries a causality-violating
instantaneous component has also been recently emphasized by Yang.\cite{10}
The effect of the acausal part of $\v A_C$ vanishes identically when we
take the curl and obtain $\gradv\times\bfA_C=\gradv\times \bfA_L$. A direct calculation 
gives\cite{7} $-\partial\bfA_C/\partial t=-\gradv\phi_L- \partial\bfA_L/\partial t +
\gradv\phi_C$. The last (acausal) term cancels exactly the instantaneous
electric field $- \gradv\phi_C$ generated by $\phi_C$ and we
again obtain $-\gradv\phi_C- \partial\bfA_C/\partial t=-\gradv\phi_L-
\partial\bfA_L/\partial t $. This expression has also been recently
demonstrated in Ref.~\onlinecite{11} using a different approach (see
Eq.~(29) in Ref.~\onlinecite{11}). In other words, the explicit presence of
an acausal term in Eq.~\eqref{eq:davis} when it is written in terms of the
Coulomb gauge potentials is irrelevant because such a term is
always canceled, which means that causality is never effectively lost.

Similar conclusions can be drawn when Eq.~\eqref{eq:davis} is expressed in terms of the Kirchhoff
gauge potentials $\phi_K$ and $\v A_K$.\cite{8} In this case the potential
$\phi_K$ propagates with the imaginary speed $ic$ and generates the imaginary field $-\gradv\phi_K$.
The Kirchhoff gauge vector potential $\v A_K$ 
contains three parts: one is causal (retarded), one is imaginary, and the remaining one mixes imaginary and retarded
contributions (see Eq.~(42) in Ref.~\onlinecite{8}). The effect of the imaginary terms in the last two parts 
vanishes identically when we take the curl and obtain
$\gradv\times\bfA_K=\gradv\times \bfA_L$. A direct calculation
gives\cite{8} $-\partial\bfA_K/\partial t=-\gradv\phi_L- \partial\bfA_L/\partial t +
\gradv\phi_K$. The last term cancels exactly the imaginary field
$-\gradv\phi_K$ and we again obtain $-\gradv\phi_K- \partial\bfA_K/\partial t=-\gradv\phi_L-
\partial\bfA_L/\partial t$. The explicit presence of an imaginary term 
in Eq.~\eqref{eq:davis} when it is written in terms of the
Kirchhoff gauge potentials is irrelevant because such a term is
always canceled, which means that causality is never effectively lost.

In the same sense that the Helmholtz theorem is considered as the
mathematical foundation of electrostatics and magnetostatics, the
generalized Helmholtz theorem can be considered as the mathematical
foundation of electromagnetism. I advocate the use of both formulations of
the generalized Helmholtz theorem\cite{12} in courses of electromagnetism
and invite instructors to decide which formulation they find more useful.


\begin{thebibliography}{99}
\bibitem{12} The author thanks V. Hnizdo for drawing his attention to the
fact that both formulations of the generalized theorem are equivalent.
\end{thebibliography}
\end{document}